\documentclass[12pt]{article}
\usepackage{sbc-template}

\usepackage[T1]{fontenc}
\usepackage[utf8]{inputenc}

\usepackage[utf8]{inputenc}
\usepackage[nomessages]{fp}
\usepackage{graphicx,url}
\usepackage{booktabs}

\usepackage{sidecap}

\usepackage{makecell}

\usepackage{amssymb}
\usepackage{cancel}

\usepackage{url}
\usepackage{siunitx}
\usepackage[flushleft]{threeparttable}
\usepackage{multirow}
\usepackage{makecell}
\usepackage{array,xstring}
\usepackage{numprint}

\usepackage{verbatim}

\usepackage{array, tabularx, boldline}
\usepackage[inline]{enumitem}

\usepackage[table,xcdraw]{xcolor}

\newcommand{\Title}{A Multi-Strategy Approach to Overcoming Bias in Community Detection Evaluation}

\newcommand{\citep}[1]{\cite{{#1}}}

\newcommand{\iAPS}{{APS}}
\newcommand{\iARXIV}{{arXiv}}

\newcommand{\iPUBMED}{{PubMed}}

\newcommand{\iHS}{{High School}}

\newcommand{\nocitedb}[1]{}

\definecolor{darkgreen}{rgb}{0.0, 0.2, 0.13}
\definecolor{red}{rgb}{1, 0, 0}
\definecolor{max}{rgb}{0, 0, 0}
\definecolor{min}{rgb}{0, 0, 0}

\newcommand{\al}[1]{{\textcolor{black}{#1}}}
\title{\Title}

\author{
Jeancarlo C. Leão\inst{1}\and
Alberto H. F. Laender\inst{2}\and 
Pedro O. S. {Vaz de Melo}\inst{2}
}

\address{
Instituto Federal do Norte de Minas (IFNMG) \\
39600-000 -- Araçuaí -- MG -- Brazil.
\nextinstitute
Departamento de Ciência da Computação \\
Universidade Federal de Minas Gerais\\
31270-901 --  Belo Horizonte -- MG -- Brazil.
\email{jeancarlo.leao@ifnmg.edu.br, laender@dcc.ufmg.br, olmo@dcc.ufmg.br}
}

\begin{document} 

\maketitle

\begin{abstract}
Community detection is key to understand the structure of complex networks. However, the lack of appropriate evaluation strategies for this specific task may produce biased and incorrect results that might invalidate further analyses or applications based on such networks. In this context, the main contribution of this paper is an {approach that} supports a robust quality evaluation when detecting communities in real-world networks. In our approach, we use multiple strategies that capture distinct aspects of the communities. The conclusion on the quality of these communities is based on the consensus among the strategies adopted for the structural evaluation, as well as on the comparison with communities detected by different methods and with their existing ground truths. In this way, our approach allows one to overcome biases in network data, detection algorithms and evaluation metrics, thus providing more consistent conclusions about the quality of the detected communities. Experiments conducted with several real and synthetic networks provided results that show the effectiveness of our approach.
\end{abstract}

\section{Introduction}

The community detection problem has been much studied in the context of social networks due to its wide application in many domains, giving rise to many methods to address it~\cite{almeida2012towards,Fortunato201075,YangLeskovec2015}\nocitedb{JCLJISA2018}.  
However, one of the major challenges related to this problem is the difficulty to evaluate the detected communities with respect to the various methods proposed in the literature. Part of this difficulty lies on the fact that there is still no universally accepted definition for the concept of community~\cite{Fortunato201075}, as well as for what we understand as being the quality of a community~\cite{Hric2014}. 
In general, this evaluation is done without explicitly dealing with bias on data, methods and metrics, which may lead to inconsistent results.

\begin{figure*}[t]
     \begin{center}            
\includegraphics[width=0.8\textwidth]{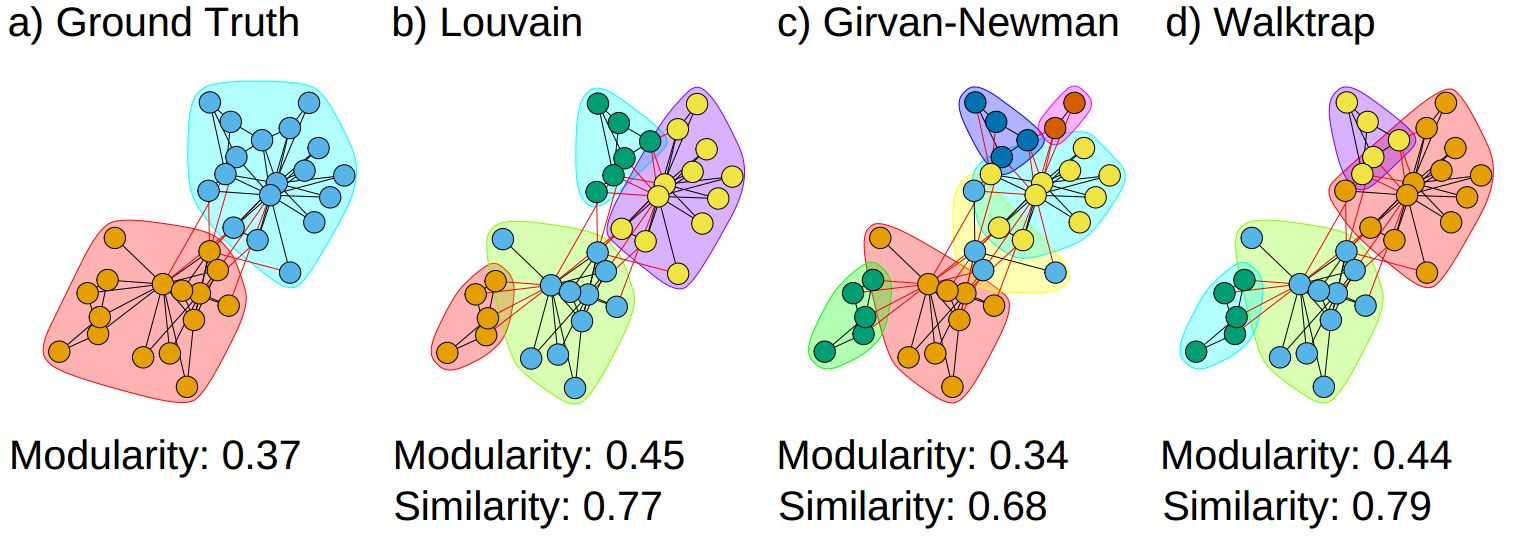}
    \end{center}
    \caption{Example of how bias can affect community  detection in social networks.}
    \label{graphic:ToyNetwork}
\end{figure*}

In order to illustrate this, let us consider the example shown in Figure~\ref{graphic:ToyNetwork}. Specifically, Figure~\ref{graphic:ToyNetwork}a shows a social network formed by 34 members (vertices) of a karate club interconnected by edges representing interactions between them outside the club. Originally, this network was divided into two non-overlapping communities labeled by Zachary~[1977]\nocite{zachary1977information} with 16 and 18 members, respectively, each one supervised by a specific instructor. Figure~\ref{graphic:ToyNetwork}b, on the other hand, shows the communities detected in this same network by the Louvain algorithm~\cite{blondel20081742-5468-2008-10-P10008}. 
Note that the community structure revealed by the Louvain algorithm is different from those presented in Figure~\ref{graphic:ToyNetwork}c and in Figure~\ref{graphic:ToyNetwork}d, which were respectively obtained by the Girvan-Newman~\cite{NewmanGirvanPhysRevE.69.026113} and Walktrap~\cite{Pons2005} algorithms, both of them considered very effective. In addition, the value of the modularity metric indicates that the communities shown in Figure~\ref{graphic:ToyNetwork}b present a better quality with respect to their modular structure (i.e., it presents the highest modularity value). On the other hand, when comparing the detected communities with the ground truth (Figure \ref{graphic:ToyNetwork}a) using the Rand Index similarity metric~\cite{Rand-JASA1971}, it indicates that the network in Figure~\ref{graphic:ToyNetwork}d is the best one, since its communities are more similar to the ones shown in Figure~\ref{graphic:ToyNetwork}a, even though there is not a perfect match. Finally, due to some specific bias in the original network, such as sporadic interactions \cite{JCLJISA2018}, the Girvan-Newman algorithm has not been able to identify good communities (Figure~\ref{graphic:ToyNetwork}c), as shown by the values of the two metrics considered. 

Thus, in face of these pieces of evidence pointing to opposite directions with respect to the quality of the communities in our example, a question arises on which one presents the best structure and what causes this divergence. 
One possible explanation to this fact is the lack of a comprehensive evaluation approach that considers multiple strategies and allows one to find out which one provides the best interpretation. More importantly, as we detail later, due to their own biases
it is not always possible to find a consensus among different metrics and community detection methods on which community structure presents the highest quality. This requires a cross-checking approach 
involving more than two distinct evaluation strategies in order to indicate a consensus and estimate a possible bias with respect to the quality of the revealed communities. 

A method that is generally employed to increase data reliability and validity is triangulation\footnote{According to O'Donoghue and Punch~[2003]\nocite{ODonoghue&Punch2003}, triangulation is a ``method of cross-checking data from multiple sources to search for regularities in the research data.''}, which consists in using multiple methods to test a same hypothesis~\cite{JCLLD2018}. 
Based on this idea, the main contribution of this paper is a robust approach for community quality evaluation that allows one to obtain results less prone to bias {when detecting communities in synthetic and real networks}. 

Thus, given a network, its set of ground truth communities and a set of its communities to be evaluated, our approach allows one to overcome biases in network data, detection methods and evaluation metrics by using distinct evaluation strategies when analyzing the quality of such communities.
For this, each strategy must strongly highlight a distinct aspect of a community's quality by considering multiple metrics, detection methods and distinct datasets.
For example, in Figure~1 the structural and functional aspects of the communities are represented, respectively, by their modularity and similarity with the respective ground truths. 
Notice that, for our purpose, the choice of the best metrics, detection methods and datasets
is not important, since we are not trying to identify the best existing community, but the best one among those being compared. 

The rest of this paper is organized as follows. Section~\ref{section:fundamentos} briefly reviews related work. Section~\ref{section:approach} describes our approach to community detection evaluation. Then, Section~\ref{section:results} analyzes the experimental results obtained by applying our proposed approach to real and simulated networks. 
Finally, Section~\ref{section:conclusion} presents our conclusions and some considerations on future work.

\section{Related Work}
\label{section:fundamentos}

\begin{table}[!tb]
\begin{center}
\caption{{Methods for community detection.}}
\label{table:communitydetectionalgorithms}
\resizebox{0.8\textwidth}{!}{
\begin{threeparttable}
    {
\begin{tabular}{@{}llcl@{}}
Main Method & Algorithm & $\xi$ & References                                                        \\ \midrule
\multirow{3}{*}{\begin{tabular}[c]{@{}l@{}}Modularity\\maximization\end{tabular}}              & Louvain Modularity (LM)                                         & D     & \cite{blondel20081742-5468-2008-10-P10008} \\ 
                                                      & \begin{tabular}[c]{@{}l@{}}Greedy Optimization\\ of Modularity (GM)\end{tabular} & D     & \cite{Clauset2004PhysRevE.70.066111}       \\ 
                                                      & Leading Eigenvector (LE)                                        & D     & \cite{Newman2006}                          \\ \midrule
Dynamic node labeling                                 & Label Propagation (LP)                                          & N     & \cite{Raghavan2007}                        \\ \midrule
\begin{tabular}[c]{@{}l@{}}Removal of edges\\ between communities\end{tabular}                  & Girvan–Newman (GN)                                           & D     & \cite{NewmanGirvanPhysRevE.69.026113}      \\ \midrule
\multirow{2}{*}{\begin{tabular}[c]{@{}l@{}}Node closeness given\\by random walks\end{tabular}} & Walktrap (WT)                                                   & N     & \cite{Pons2005}                            \\
                                                      & Infomap (IM)                                                    & N     & \cite{10.1371/journal.pone.0018209}        \\ \bottomrule
\end{tabular}
\par
\begin{tablenotes}
\item $\xi$: State model (D-Determinístic/N-Non determinístic).
\vspace{-2mm}
\end{tablenotes}
}
\end{threeparttable}}
\end{center}
\end{table}

Although community detection has become one of the most popular and best-studied research topics in network science~\cite{Zhao2017}, the problem of validating the quality of a community derived from a real network has not received the due attention, since there is no consensus on what is meant by a good community. For example, the methods listed in Table~\ref{table:communitydetectionalgorithms} usually extract distinct communities from a given network, which are usually considered of good quality by different metrics.

In this context, there is no best metric to assess the quality of a community~\cite{almeida2012towards}. Moreover, community detection algorithms are usually evaluated by correlated metrics or by the same metrics used by their optimization function, such as modularity \cite{Fortunato201075,YangLeskovec2015}. 
Thus, existing work usually considers only  specific aspects to assess the quality of a community, for example by measuring the structure derived from its connectivity \cite{NewmanGirvanPhysRevE.69.026113}, comparing its similarity with a ground truth community \cite{Peele1602548} or performing a comparison with a good baseline \cite{Hric2014}.

Regarding the structural aspect, popular quality metrics present strong bias when applied to networks with different sizes or number of clusters~\cite{almeida2012towards,coscia2011classification,Pons2005}. In particular cases, it is possible to assess the functional aspect of a detected community by comparing it with its respective ground truths~\cite{Hric2014,Peele1602548,zaki2014dataminingbook}.
For Fortunato et al.~[2010]\nocite{Fortunato201075}, this kind of evaluation involves the definition of a criterion to establish how ``similar'' is a community provided by an algorithm with respect to the ground truth. To address this, the authors adopt some specific indexes such  as \textit{Rand Index} and \textit{Normalized Mutual Information}. 

In addition, community detection algorithms are sensitive to different community structures, topologies or instances of a network \cite{coscia2011classification}.
In this context, different approaches have been proposed with the aim of reducing the effect of biases and improving the detection of communities. For instance, Lancichinetti et al. [2012] \nocite{Lancichinetti2012} show how to combine the communities obtained from various detection methods into a consensual one, statistically more stable and with a better structure.
In a previous work, Rocha et al. [2017]\nocite{PhysRevE.96.052302} described how the representation of real temporal interactions can result in biased data.  
More recently, Leão et al. [2018]\nocitedb{JCLJISA2018} 
proposed a solution to the biased data problem by directly removing noisy produced by sporadic relationships\footnote{In the history of interactions of a social network there are those that represent a strong relationship between two people in a community (e.g., a teacher and a student in a school) and others, result of chance, that represent interactions between people from different communities and most likely will not occur in the future (e.g., a phone call from a telemarketer) \cite{JCLJISA2018}.
} found in a social network. In addition, they also showed that this kind of noise may cause errors when detecting communities. 

To the best of our knowledge, the closest work to ours is the one by  Yang and Leskovec~[2015]\nocite{YangLeskovec2015}. 
In their work, they use the correlation between distinct community definitions to evaluate their structure in large social networks. On the other hand, Lancichinetti and Fortunato [2012]~\nocite{Lancichinetti2012} seek consensus only on the structural aspect of the communities. In both works, the authors evaluate the quality of a community without aiming at a consensus involving distinct aspects or addressing any bias. 

Thus, by analyzing the above works, we have not been able to identify any approach that deals with different types of bias for assessing the quality of a community. 
Moreover, differently from our work, the above ones do not  provide a systematic and consistent strategy to produce a robust conclusion about a community's quality.

\section{Proposed Approach}
\label{section:approach}
\label{subsec:evalImprov}

\begin{figure*}[!t]
     \begin{center}            
\includegraphics[width=1\textwidth]{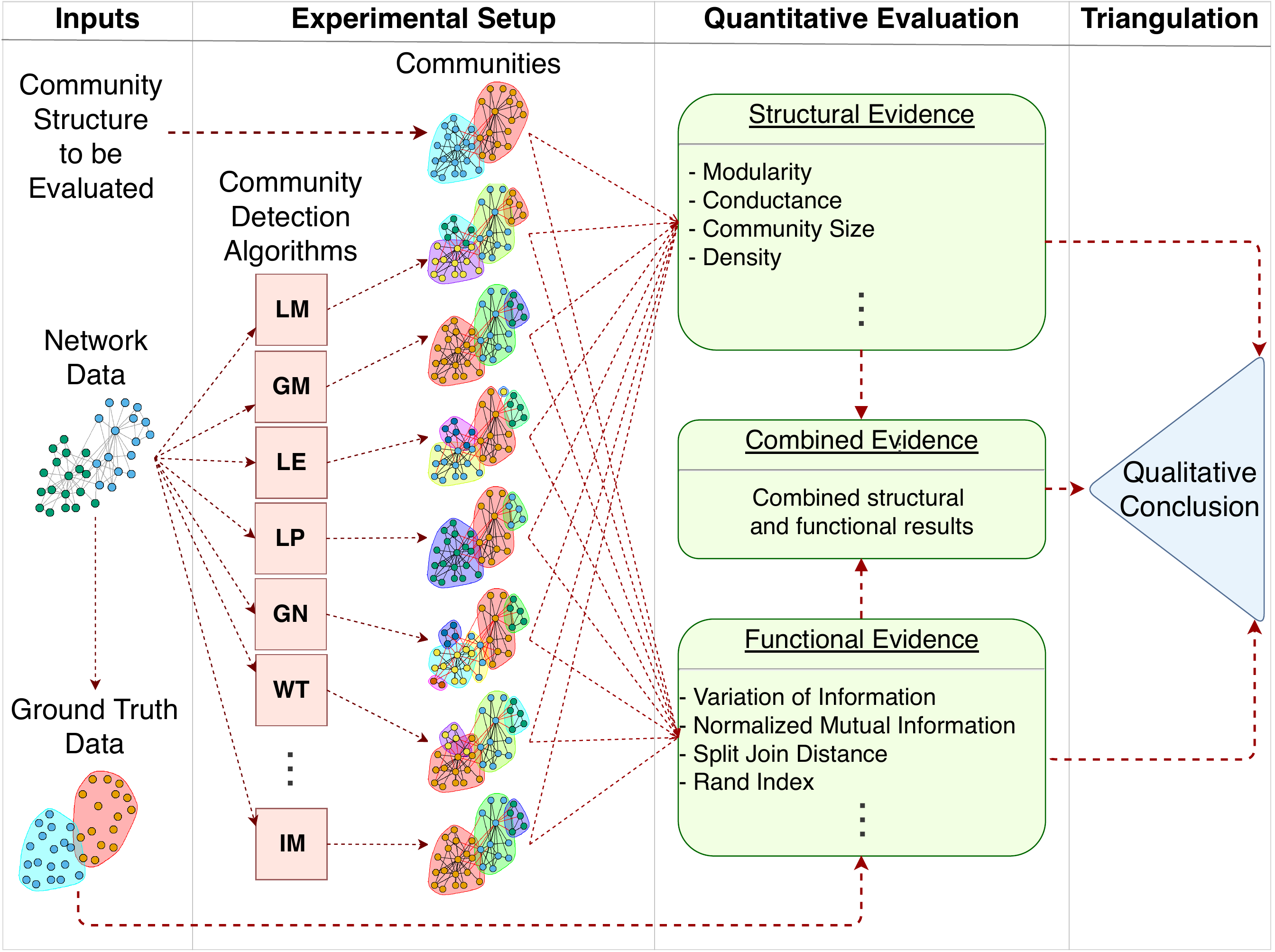}
    \end{center}
    \caption{Overview of the proposed approach to community detection evaluation.}
    \vspace{-2mm}
    \label{fig:approach}
\end{figure*}

Figure~\ref{fig:approach} summarizes our approach. First, we provide as input a network, its ground truth communities and the set of its communities that we wish to evaluate. Next, in addition to a set of ground truth communities, we consider as further evidence the communities detected by multiple algorithms, for example, those listed in Table~1. Then, in the quantitative evaluation step, all communities are assessed by multiple structural and functional metrics, and then compared to each other to provide a set of combined evidence. Finally, we group the results produced by each algorithm in a new set of pieces of evidence in order to highlight structural and functional aspects related to the quality of each community, and compare them with those of the communities obtained by the other algorithms, as described next.

\pagebreak
By using structural metrics, we quantify how much the connectivity of specific sets of nodes in the network expresses structural characteristics that are typical of real-world communities~\cite{YangLeskovec2015}. For this, we take into account multiple pieces of evidence on the quality of a community expressed by metrics such as modularity, conductance and density~\cite{NewmanGirvanPhysRevE.69.026113,YangLeskovec2015}. We also use specific statistics, such as the number and size of the detected communities, the variance of these values and network measures to help  analyzing the results. 
In addition, we consider similarity (or functional) metrics such as \textit{Variation of Information}~(VI), \textit{Normalized Mutual Information}~(NMI), \textit{Split Join Distance}~(SJD) and \textit{Rand Index}~(RI)~to provide some functional evidence.

We then combine structural and functional measures to compare the quality of the communities obtained by different community detection algorithms applied to the same network. 
For this, we assess the agreement between these measures by using the standard one-dimensional Euclidean distance. By crossing all types of evidence considered (structural, functional and combined), we capture distinct aspects of the communities' quality and conclude on the quality of the their structures by means of the consensus among all pieces of evidence. 
In addition, we check the effect of bias in data by controlling its source.
More specifically, to minimize and estimate the effects of bias caused by noisy data, we filter the networks by using the framework proposed in our previous work~\cite{JCLJISA2018}. 
Note that here ``data bias'' is any error generated by a community detection algorithm that might be associated with noise in the network. 

\section{Experimental Results}
\label{sec:results}
\label{section:results}

To evaluate our proposed approach, we run a series of experiments to assess the structural and functional aspects of the communities derived from five networks by applying a combination of seven algorithms based on state-of-the-art detection methods. 
Note that in these experiments we analyze the communities generated by each algorithm separately, considering the other ones as their baselines. In addition, experiments involving non-deterministic algorithms (see Table~1) were performed several times to ensure the reliability of the results.

\ifx\NOSHOWFIG\defined
\begin{table}
\centering
\caption{{Caracterization of the networks.}}
\label{table:networks_statistics}

\resizebox{0.75\textwidth}{!}{
\begin{threeparttable}
    {
\begin{tabular}{@{}c|lrrrrrrr@{}}   
Appl. Domain & \multicolumn{1}{c}{{Network}} & \multicolumn{1}{c}{$|V|$} & \multicolumn{1}{c}{$|E|$} & \multicolumn{1}{c}{$\Delta$} & \multicolumn{1}{c}{$D$} & \multicolumn{1}{c}{\emph{CC}} \\ \midrule
\multirow{3}{*}{\begin{tabular}[c]{@{}c@{}}Scientific \\ Collaboration\end{tabular}} & \iAPS  & 181k & 852k  & 305 & 0.5 & 0.33  \\
 & \iPUBMED & 444k & 5.5M & 4869 & 0.6 & 0.36 \\
 & {arXiv} & 33k & 180k & 424 & 3.3 & -  \\
\hline
\begin{tabular}[c]{@{}c@{}}Diseace \\ Propagation\end{tabular} & {High~School} & 327 &  5818 & 87 & 1.1k & 0.44 \\
\hline
Simulated Nets & {Sinthetic} & $\approx$1k & $\approx$13k & $\approx$78 & $\approx$267 & $\approx$0.29 \\
\bottomrule
\end{tabular}
\par
\begin{tablenotes}
\small{\item $|V|$: set of vertices; $|E|$: set of edges; $\Delta$: max degree; $D$: density (x${10^{-4}}$); $CC$: cluster coeficient. The min degree is 1 in all networks. 
}
\end{tablenotes}
}
\end{threeparttable}}
\end{table}

\subsection{Networks}

Initially, we modeled as aggregate edge graphs the scientific collaboration networks (here identified by their respective datasets, namely APS, PubMed and arXiv)\footnote{Datasets obtained from http://homepages.dcc.ufmg.br/\~{}mirella/projs/apoena/. 
APS: co-authorship network of members of the American Physical Society; PubMed: co-authorship network derived from scientific articles available on MEDLINE; arXiv: co-authorship network derived from scientific articles obtained from https://www.kaggle.com/neelshah18/arxivdataset/} and the contact network of secondary school students\footnote{Datasets obtained from http://www.sociopatterns.org/datasets/.}, which were used in previous works by  Gemmetto et al. [2014]\nocite{Gemmetto2014} and~Le\~ao~et al. [2018]\nocite{JCLLD2018}, 
respectively.
Table~\ref{table:networks_statistics} presents a general characterization of these networks. 

Notice that these networks represent distinct social relationships.
Thus, in the scientific collaboration networks, vertices represent researchers and there is an edge connecting two researchers if they are coauthors of a same article. In the contact network, vertices represent members of a school (for example, students or teachers) and there is an edge between two individuals if they are close to each other. 
We also used synthetic networks for which we have created  their respective ground truth communities. 
These synthetic simulated networks are based on the GRM model \cite{grm}, 
which allows the representation of mobility networks with group (community) characteristics.  

\begin{figure}[h]
    \centering
    \includegraphics[width=0.5\textwidth]{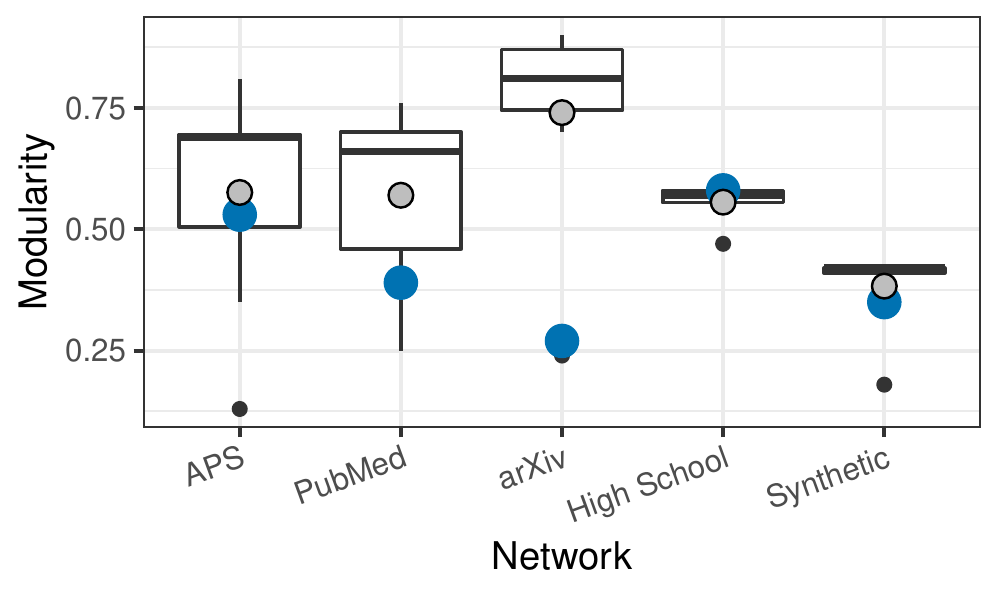}\includegraphics[width=0.5\textwidth]{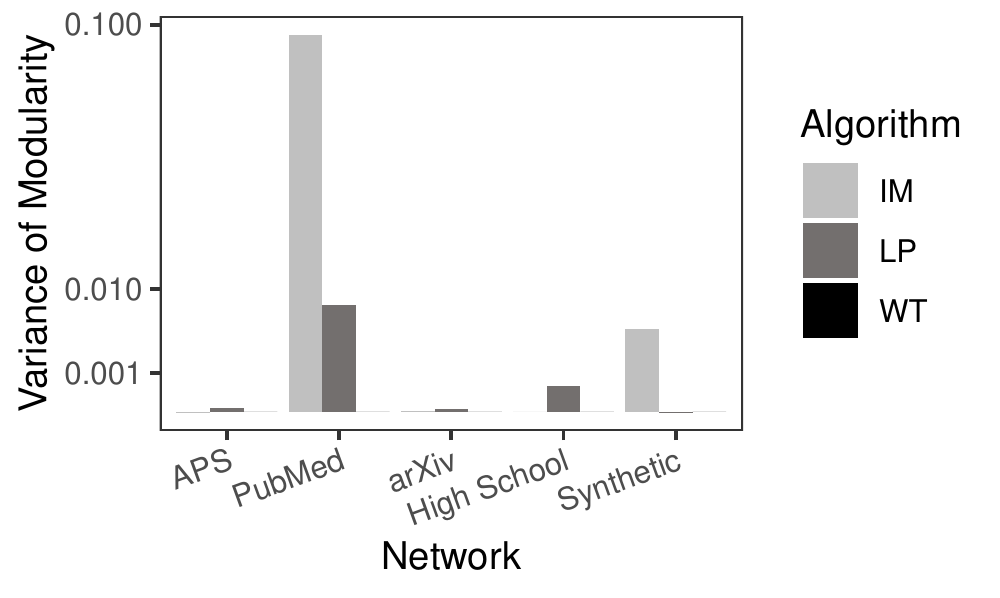}
    \caption{Modularity values for communities detected by all considered algorithms (boxplot) and for the ground truth (blue dot). $Var(Modularity)$: Variance of the modularity between replications of the detection experiments by non-deterministic algorithms (deterministic algorithms have zero variance and therefore are not presented in the right graph).}
    \vspace{-2mm}
    \label{graphic:modularityagreement}    
\end{figure}

\subsection{Evidence Considered}

The combination of functional and structural evidence in our experiments allowed us to corroborate the quality of the ground truths as well as of the communities detected in all networks. This also made it possible to indicate the algorithm that identified the best communities on the networks. For this, we first analyzed the results of each strategy individually, providing hypotheses about the quality of the communities. Then, we combined these results, verifying the consensus among the communities. 
In this way, we verified which hypotheses were refuted, as well as the biases identified. 

\begin{figure}[!h]
\includegraphics[width=\textwidth]{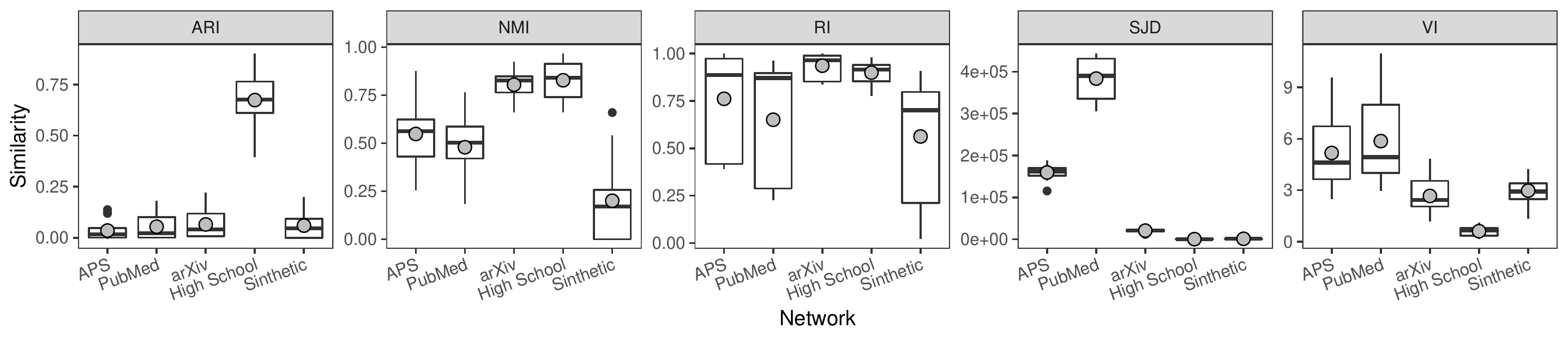}
\vspace{-4mm}
\caption{Similarity values between communities detected by different algorithms and measured by different metrics. Note that, especially for the metrics VI and SJD, the lower their values, the greater the similarity indicated.
}
\label{graphic:similarity_alg}
\end{figure}
\vspace{-5mm}

\subsubsection{Identifying the Best Communities}

First, we analyze the structure of the communities detected by the different algorithms. Here, we note that the communities derived from the High School and arXiv networks have the best well defined characteristics. 
For this, we consider the following evidence:
high average modularity (Figure~\ref{graphic:modularityagreement}, left), greater consensus on the structure of the communities (interquartile of the similarity between them, presented in Figure~\ref{graphic:similarity_alg}), greater confidence of the modularity  value obtained in different experiments with the same non-deterministic algorithm (Figure~\ref{graphic:modularityagreement}, right) and small variation in the number of communities detected by
these algorithms (Figure~\ref{graphic:communitycount}). However, as we shall see below, although such pieces of evidence indicate that the communities from these two networks have the same characteristics, we have not come to the same conclusion about their quality. 

\begin{figure}[h]
\includegraphics[width=\textwidth]{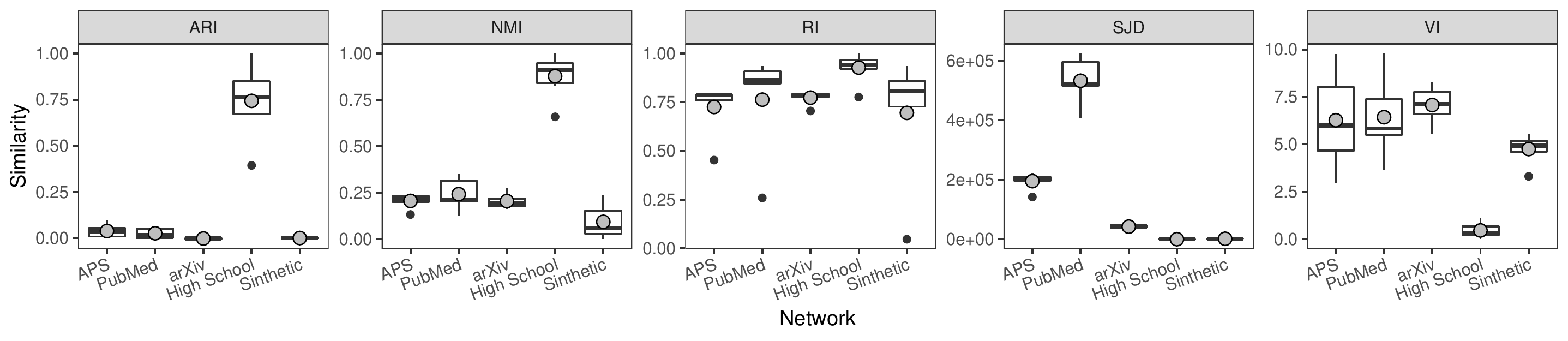}
\caption{Values of the similarity between the ground truth and the communities detected by distinct algorithms and expressed by different metrics.
}
\label{graphic:similarity_gt}
\end{figure}

From a functional viewpoint, unlike the High School network, in the arXiv one there is no convergence of evidence to confirm the quality of its communities when compared with the ground truth (Figure~\ref{graphic:similarity_gt}). This can be considered as a disagreement with respect to the structural aspect when we compare the distance between the modularity values of the  arXiv network with those of the other networks. 

Note in Figure~\ref{graphic:modularityagreement} (left), for example, that there is a large difference between the modularity values of the ground truths and those estimated for the detected communities. In addition, according to Figure~\ref{graphic:communitycount}, the number of communities in the ground truths is far from the number of communities actually detected in the networks. Therefore, the strength of this initial evidence has led us to the conviction that the communities detected in the actual networks are the correct ones. In addition, the confidence and the structural evidence that strongly disagree with the functional one corroborate the interpretation that the detected communities are the real ones and not those shown by the ground truths. 

\begin{figure}[h]
    \centering
        \includegraphics[width=0.45\textwidth]{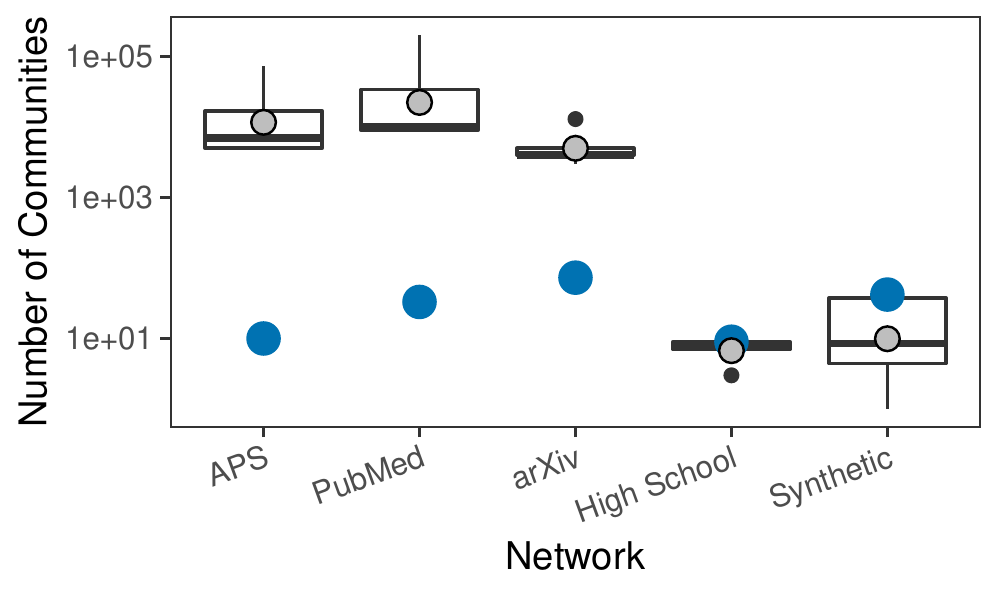}
    \caption{Number  of communities detected by all considered algorithms in each network  (boxplot) and in the respective ground truth (blue dot).
    }    
    \label{graphic:communitycount}    
\end{figure}

Note that the two sets of evidence, structural and functional, contradict each other on which communities in the arXiv network are the best ones and also on which evidence in the first set is the strongest one. Based on our approach, the possibility of bias being the cause of these divergences makes it necessary to evaluate them by using a third set of evidence (on a particular aspect) to support one of the two contradictory pieces of evidence. For this, we identified the predominant source of bias and showed that it interferes with the final conclusion.

It occurs that in the arXiv network the bias caused by the detection methods does not considerably interfere in their results, since the communities suggested by them are structurally similar. In addition, this was evidenced in this network by all structural metrics considered, whose values corroborate a high-quality community structure, as already shown in Figures \ref{graphic:modularityagreement}, \ref{graphic:similarity_alg} and  \ref{graphic:communitycount}.
Thus, we hypothesized that the interference bias is predominantly in the data, provoking a disagreement between structural and functional evidence, as well as the identification of false communities of high quality.

For this, we first analyze the ground truth communities of the arXiv network and then 
consider the meaning of these communities in this network, i.e., they are groups of researchers that publish together and predominantly in the same area of knowledge. However, it should be noted that this definition is not absolute since there may be a multidisciplinary community with sporadic co-authorships or a community of researchers {that work in the same area, but do not collaborate with each other.}  In both cases, ground truth communities are not very well captured by detection methods that rely on network connectivity.
Thus, we consider the hypothesis that the bias that obscures the real community structure of the arXiv network is a consequence of the existence of edges and nodes that represent, respectively, {sporadic collaborations and researchers that work in the same knowledge area, but} do not significantly interact with each other.

Then, to test our hypothesis, we run the following bias control experiment: we removed such skewed edges and nodes using the filtering framework proposed by Leão et al. [2018]\nocite{JCLJISA2018} and then applied the same detection algorithm on the filtered version of that network\footnote{The generated datasets are available by request at {http://cnet.jcloud.net.br} repository.}.
This time we obtained new communities in which the convergence between the structural and functional metrics occurred, as indicated by its greater similarity with the ground truth communities, and having a better structural aspect, as indicated by all metrics used for this purpose. 
This way, we have been able to identify that the most significant source of bias in the \iARXIV~network was its data, which shows that considering few pieces of evidence can lead to {apparently convincing results, but unreliable.}

In the APS and PubMed networks, data bias is also the main source of divergence between the respective ground truths and the detected communities. However, two characteristics of these two ground truths are among those that most interfere in the quality of their respective communities: the high overlap among the communities [Leão et al. 2018] and the large number of communities formed by multiple components (see Figure~\ref{graphic:communitycount}). 
This shows how pieces of structural evidence, such as those obtained by modularity, are insufficient to characterize this disagreement, even though the modularity of the ground truth and the communities detected in these networks have relatively high values and are close to each other. We also verify that the bias in the data caused by sporadic relationships does not  considerably interfere in the detection of the communities since there was no significant improvement after filtering the networks. On the other hand, as shown in Figure \ref{graphic:similarity_alg}, there was little consensus among the communities detected in these networks by the different algorithms.

In this context, in addition to identifying significant bias in the structure of the arXiv and High School networks, this phenomenon was also verified in a smaller scale among the detection methods. For example, as demonstrated for the arXiv network, its consensual community, despite its high structural quality, presents results that are considerably skewed.
On the other hand, in the \iPUBMED~network and more clearly in the \iAPS~network, the average quality of the communities detected by different algorithms stands out when compared with the ground truths. 
In the synthetic networks and in the \iHS~one bias had no significant interference on the convergence of the structural, functional and combined pieces of evidence of the communities' quality. 

\subsubsection{Best Detection Algorithms}

 Although the most modular communities are those detected by the Louvain algorithm (upper bounds shown in Figure~3, left), the modularity values of the communities detected by the Infomap algorithm are generally closer (there is a greater agreement) to those of the ground truth. 
 In addition, Infomap provided a number of cases in which there was an agreement between the modularity of the detected communities and that of the ground truth. 
 On the other hand, these same metrics achieved smaller values for the Louvain algorithm. In addition, the modularity of the communities extracted by different algorithms and that of the functional communities have considerably varied for most networks. 
We also verified how different community detection algorithms agree with each other and with the network ground truths with respect to their communities. Despite the variation in the structure of the detected communities (Figure 4), we verified a higher consensus among them than with the ground truth communities, as shown in Figure 5.

\begin{table}[]
\caption{Best detection algorithm according to distinct experiments.}
\begin{center}
\resizebox{1\textwidth}{!}{
\begin{threeparttable}
    {
\begin{tabular}{@{}llllllll@{}}
Best Algorithm  & LM            & GM  & LE          & LP             & WT               & IM                    \\ \midrule
Metrics & ARI, SJD      & ARI & SJD*, VI*     & RI, SJD, VI    & ARI, RI          & ARI, NMI*, RI, SJD, VI\\ 
Networks & PubMed, arXiv & APS & APS, PubMed & APS, Sinth. & arXiv, Sinth. & All                  
\\ \bottomrule
\end{tabular}
\par
\begin{tablenotes}
\item *Metrics with the best overall value.
\end{tablenotes}
}
\end{threeparttable}}
\end{center}
\label{table:bestmethod}
\end{table}

In addition to obtaining a consensus among different algorithms, our approach also identified some algorithms with distinct behavior, such as Infomap, that detected less modular communities, but in general more similar to their ground truths. Despite such divergences among the strategies, most pieces of evidence indicate the Louvain algorithm as the least biased and the one that obtained the best values for most of the structural metrics, particularly modularity. 
We also identified some algorithms that presented the best score when considering a specific metric. This is the case of the Louvain  algorithm (LM) for modularity, and of the Infomap (IM) and Leading Eigenvector (LE) algorithms for the similarity metrics Normalized Mutual Information (NMI), and Split Join Distance (SJD) and Variation of Information (VI), respectively (see Table~3).
Notice that our proposed approach is able to analyze distinct alternative solutions for the task at the hand, thus being able to identify those that provide the best trade-off.  

\section{Conclusions and Future Work}
\label{section:conclusion}
The main contribution of this paper is an approach to identify and reduce {effect of biases} when assessing the quality of communities detected by distinct algorithms. Specifically, we use multiple and diversified measurement strategies designed to capture different aspects of the quality of a community structure. 
{For its evaluation, we carried a set of experiments using five networks (four real ones and one synthetic) and compared the results obtained by seven community detection algorithms considered the state-of-the-art in the area.} In addition, we also evaluated {the quality of the communities by using different strategies.}
In this context, our evaluation evidentiated the bias of each strategy, thus providing some consensus among them.

\al{By doing so,} we were able to sustain our hypothesis by showing that the quality evaluation of communities detected from a network must be supported by multiple pieces of evidence. That is, given the discrepancy between the quality indicated by distinct evaluation strategies, {we evidentiate} that the use of a single quality metric, be it structural or functional, makes the results biased and unreliable. \al{On the other hand}, our multi-strategy  evaluation approach made it possible to explain extreme values for some of the metrics considered. For example, we were able to verify the existence of bias in modularity metrics, some ground truths and network data, and some detection algorithms. 

A current limitation of our proposed approach is the use of a predefined set of evaluation metrics and community detection algorithms. However, this limitation can be easily overcome by providing a configurable platform in which such features could be defined according to specific characteristics of the networks being considered. 
Thus, as future work, we intend to conduct a study to characterize the diversity of algorithms and metrics usually used for community detection, in order to provide insights for improving our approach.
Finally, it is worth noting that the approach proposed in this paper can be adapted to other tasks besides community detection. Thus, another line of future work could be, for example, adapting this approach to assess the task of link prediction in social networks in order to provide more robust results.

\section*{Acknowledgements}

Work supported by project MASWeb (FAPEMIG/PRONEX grant APQ-01400-14) and by the authors' individual grants from CNPq and FAPEMIG. Particularly, the first author would like to thank LBD/UFMG, JCLoud.net.br and LabSiCCx - Laboratório de Sistemas Computacionais Complexos (PROPPI/IFNMG, project Nr. 209/2019) for the infrastructure provided.

\bibliographystyle{sbc}
\bibliography{sbc-template}

\end{document}